\documentclass[prl, twocolumn,preprintnumbers,amsmath,amssymb,aps,pra,superscriptaddress]{revtex4}

\usepackage{mathrsfs}
\usepackage{graphicx}
\usepackage{dcolumn}
\usepackage{amsmath}
\usepackage{epsfig}
\usepackage{subfigure}
\usepackage{color}
\usepackage{epstopdf}
\usepackage{framed}
\usepackage{float}

\begin{document}

\date{\today}

\title{Floodlight Quantum Key Distribution: Demonstrating a New Framework for High-Rate Secure Communication}%
\author{Zheshen Zhang}
\email{zszhang@mit.edu}
\affiliation{Research Laboratory of Electronics, Massachusetts Institute of Technology,
Cambridge, Massachusetts 02139, USA}%
\author{Quntao Zhuang}
\affiliation{Research Laboratory of Electronics, Massachusetts Institute of Technology,
Cambridge, Massachusetts 02139, USA}%
\affiliation{Department of Physics, Massachusetts Institute of Technology, Cambridge, Massachusetts 02139, USA}
\author{Franco N. C. Wong}
\author{Jeffrey H. Shapiro}
\affiliation{Research Laboratory of Electronics, Massachusetts Institute of Technology,
Cambridge, Massachusetts 02139, USA}%

\begin{abstract}
Floodlight quantum key distribution (FL-QKD) is a radically different QKD paradigm that can achieve Gbit/s secret-key rates over metropolitan area distances without multiplexing [Phys.\,Rev.\,A {\bf 94,} 012322 (2016)]. It is a two-way protocol that transmits many photons per bit duration and employs a high-gain optical amplifier, neither of which can be utilized by existing QKD protocols, to mitigate channel loss. FL-QKD uses an optical bandwidth that is substantially larger than the modulation rate and performs decoding with a unique broadband homodyne receiver.  Essential to FL-QKD is Alice's injection of photons from a photon-pair source---in addition to the light used for key generation---into the light she sends to Bob.   This injection enables Alice and Bob to quantify Eve's intrusion and thus secure FL-QKD against collective attacks. Our proof-of-concept experiment included 10\,dB propagation loss---equivalent to 50\,km of low-loss fiber---and achieved a 55\,Mbit/s secret-key rate (SKR) for a 100\,Mbit/s modulation rate, as compared to the state-of-the-art system's 1\,Mbit/s SKR for a 1\,Gbit/s modulation rate [Opt.\,Express {\bf 21,} 24550--24565 (2013)], representing $\sim$500-fold and $\sim$50-fold improvements in secret-key efficiency (SKE) (bits per channel use) and SKR (bits per second), respectively.
\end{abstract}


\maketitle

Quantum key distribution (QKD) \cite{Bennett1984,Ekert1991,Grosshans2002,Lucamarini2013,Comandar2014,Korzh2015,Huang2015} enables two remote users (Alice and Bob) to create a shared secret key with unconditional security. Using that key as a one-time pad, they can then communicate with information-theoretic security. Unfortunately, propagation loss incurred in long-distance transmission has kept the secret-key rates (SKRs) of state-of-the-art QKD systems far below the Gbit/s rates needed for their widespread deployment.  The universal upper limit on a QKD system's secret-key rate \cite{Takeoka2014,Pirandola2015,Wilde2016} is $-\log_{2}(1-\kappa)$ bits per optical mode \cite{Pirandola2015}, where $\kappa$ is the channel transmissivity.  A  50\,km low-loss fiber has $\kappa=0.1$, implying a 0.15\,bit/mode rate limit.  Continuous-variable QKD protocols necessarily operate with one optical mode per channel use \cite{Grosshans2003,Jouguet2013,Huang2015}, while decoy-state BB84 (the predominant discrete-variable QKD protocol) takes no advantage of multiple modes per channel use \cite{Gobby2004,Takesue2007,Dixon2010,Tanaka2012,Lucamarini2013,Comandar2014,Korzh2015}.  Hence $-\log_2(1-\kappa)$ is their ultimate secret-key efficiency (SKE) in bits per channel use. High-dimensional QKD systems employ multiple modes per channel use \cite{Cerf2002,Thew2004,Zhang2008,Mower2013,Zhang2014,Lee2014,Zhong2015}, but their secret-key rates do not exceed that of ideal decoy-state BB84.   State-of-the-art QKD implementations, however, have SKEs well below this limit, e.g., the 1\,Mbit/s, 50-km-fiber demonstration from Refs.~\cite{Lucamarini2013,Comandar2014} realized $10^{-3}$\,bits per channel use.  Given current bandwidth limitations on electronics, such low SKEs preclude existing QKD protocols from attaining Gbit/s SKRs over metropolitan-area distances unless massive amounts of wavelength-division multiplexing---with their attendant complexity and cost---are employed.

In this Letter we experimentally validate a new QKD framework, called floodlight QKD (FL-QKD) \cite{Zhuang2016}, that is capable of Gbit/s SKRs over metropolitan-area distances.  FL-QKD transmits many photons per bit duration and uses a high-gain optical amplifier, neither of which can be utilized by existing QKD protocols, to mitigate channel loss. Transmitting many photons per bit duration without compromising security is possible because FL-QKD utilizes an optical bandwidth that is much greater than its modulation rate.  Furthermore, FL-QKD employs a unique broadband homodyne receiver that effectively leverages optical bandwidth without resorting to multiplexing.

\begin{figure}
\includegraphics[width=3.4in]{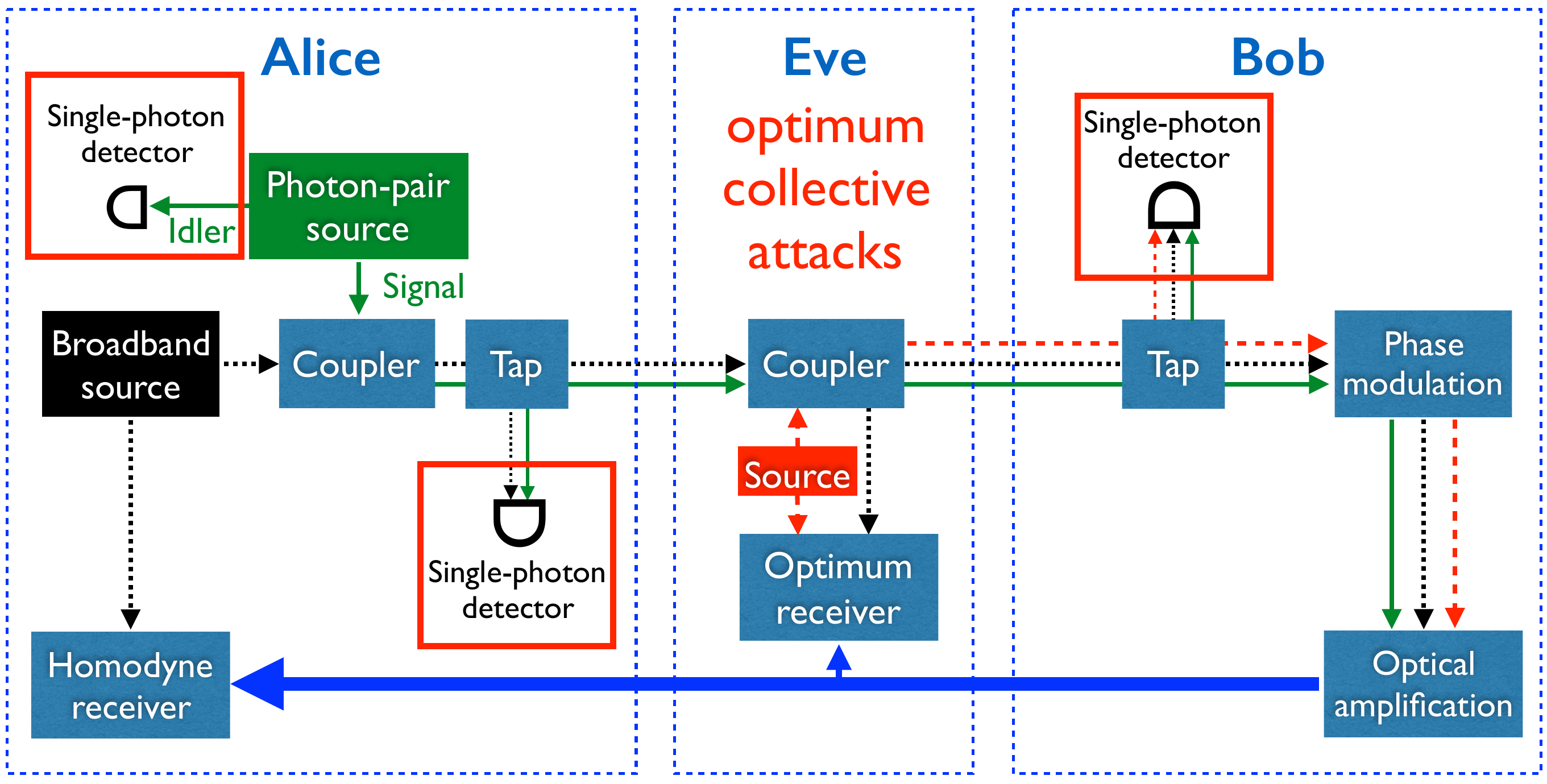}
\caption{\label{FigProtocol} (color online) Schematic of FL-QKD under Eve's optimum collective attack realized as an SPDC-injection attack. Photons generated by Alice's broadband source are marked by thin dotted lines (black); photons generated by Alice's photon-pair source are marked by thin solid lines (green); photons generated by Eve's entanglement source are marked by thin dashed lines (red); photons emitted by Bob's amplifier are marked by thick lines (blue). The channel monitoring apparatus consists of the three single-photon detectors placed in red boxes. Note that at Bob's single-photon detector only the photons originating from Alice's photon-pair source (solid line) are coincident with Alice's idler photons, whereas the photons injected by Eve (dashed line) only contribute to the noise background. As such, Eve's injection ratio $f_E$ can be determined by measuring the coincidences versus singles rates at Alice and Bob's single-photon detectors.}
\end{figure}

Figure~\ref{FigProtocol} illustrates the structure of FL-QKD. FL-QKD is a two-way multi-mode protocol in which, unlike previous two-way protocols \cite{Shapiro2009, Zhang2013,Shapiro2014,Bostrom2002, Deng2004, Pirandola2008_2WQC, Weedbrook2014}, Alice uses an amplified spontaneous emission (ASE) source whose optical bandwidth $W$ is much greater than the modulation rate $R$.  Thus in one channel use (one $T = 1/R$ bit duration) Alice's ASE source emits $M = WT \gg 1$ temporal modes of duration $1/W$.  

FL-QKD has three principal steps.  First, Alice sends a low-brightness (photons/mode $\ll 1$) portion of her ASE light to Bob, which is completely correlated with the high-brightness (photons/mode $\gg 1$) remainder that she retains for use as her homodyne receiver's local oscillator (LO).  Low-brightness light cannot be amplified or cloned with good fidelity, while the $M\gg 1$ modes per bit duration permits FL-QKD to mitigate loss in the Alice-to-Bob channel, so that on average Bob receives at least one photon per bit duration. 
  Second, Bob uses binary phase-shift keying (BPSK) to encode a bit sequence onto the light he received from Alice. He then amplifies the modulated light with a high-gain amplifier to overcome the return-path propagation loss while adding spontaneous emission noise.  Third, Alice receives the light returned from Bob and decodes his bit sequence by broadband homodyne reception after delaying her LO by that of the Alice-to-Bob-to-Alice roundtrip. 

An essential part of FL-QKD is Alice and Bob's monitoring of their quantum channel to defeat Eve's optimum collective attack \cite{Zhuang2016}. Alice combines the low-brightness ASE light she is sending to Bob with the signal output from a continuous-wave (cw) spontaneous parametric down-converter (SPDC) and taps a small fraction of the combined ASE-SPDC light for coincidence detection with the SPDC's idler output prior to sending the rest of that light to Bob. Her singles and signal-idler coincidence measurements tell her the SPDC fraction in her combined ASE-SPDC light. Similarly, Bob taps a small fraction of the light he receives for singles measurements and for coincidence measurements with Alice's idler that tell him the fraction of his received light which originated from the SPDC. Bob's tap measurements are normalized relative to Alice's tap measurements to yield $f_E$, a parameter that quantifies Eve's intrusion on the Alice-to-Bob quantum channel:  
\begin{equation}
f_E = 1-\frac{(C_{IB}-\widetilde{C}_{IB})/S_B}{(C_{IA}-\widetilde{C}_{IA})/S_A}.
\label{monitor}
\end{equation}
Here, $C_{IB}$ ($C_{IA}$) is the time-aligned coincidence rate of Bob's (Alice's) tap,  $\widetilde{C}_{IB}$ ($\widetilde{C}_{IA}$) is the time-shifted coincidence rate of Bob's (Alice's) tap that measures accidental coincidences, and $S_B$ ($S_A$) is the singles rate of Bob's (Alice's) tap. This $f_E$ measurement is calibration free \cite{Zhuang2016}, i.e., it is independent of channel loss, detector efficiencies, and source brightness.

We emphasize that FL-QKD is fundamentally a multi-mode protocol ($M \gg 1$) that makes it feasible to achieve much higher SKE and SKR than existing QKD protocols by utilizing an ASE source whose optical bandwidth greatly exceeds the modulation rate. Operationally, however, this large bandwidth disparity does not require wavelength-division multiplexing, as is the case in classical communication. Instead, a broadband homodyne receiver whose electrical bandwidth equals the modulation rate suffices. Moreover, Bob's optical amplifier, at the cost of a modest reduction in each mode's signal-to-noise ratio, compensates return-path loss and detector inefficiency. It also adds noise such that high-efficiency, shot-noise limited detection is not necessary, unlike the case for continuous-variable QKD \cite{Grosshans2002,Grosshans2003,Jouguet2013}.

In Ref.~\cite{Zhuang2016} we showed that Eve's optimum collective attack is the SPDC-injection attack \cite{SM} shown in Fig.~\ref{FigProtocol}, for which $f_E$ equals the fraction of light entering Bob's terminal that came from Eve.  The collective-attack  security analysis in Ref.~\cite{Zhuang2016} granted Eve {\em all} of the return light and an optimum collective measurement for decoding Bob's bit sequence.  A lower bound on Alice and Bob's information-rate advantage against the optimum collective attack can be obtained via 
\begin{equation}
\Delta R^{\rm LB}_{AB} = \left[\beta I_{AB} -\chi_{EB}^{\rm UB}(f_E)\right]R,
\end{equation}
where $I_{AB}$ is Alice and Bob's Shannon information, $\beta$ is their reconciliation efficiency, and $\chi_{EB}^{\rm UB}(f_E)$ is an upper bound on Eve's Holevo information when her injection fraction is $f_E$. In Eve's passive attack (no light injection) \cite{Shapiro2009}, her Holevo-information bound, $\chi_{EB}^{\rm UB}(0)$, can be determined prior to communication. Then, because Alice decodes \emph{all} the bits sent by Bob, they can perform direct secure communication by Bob's using codes developed for wiretap channels to send a message to Alice with sematic-security protection \cite{Bloch2015}. However, when Eve injects light into Bob ($f_E > 0$), her Holevo-information bound, $\chi_{EB}^{\rm UB}(f_E)$, can only be determined after channel monitoring. In this case, Alice and Bob can only perform key distribution and Bob must therefore transmit a random bit sequence.

\begin{figure*}
\includegraphics[width=\textwidth]{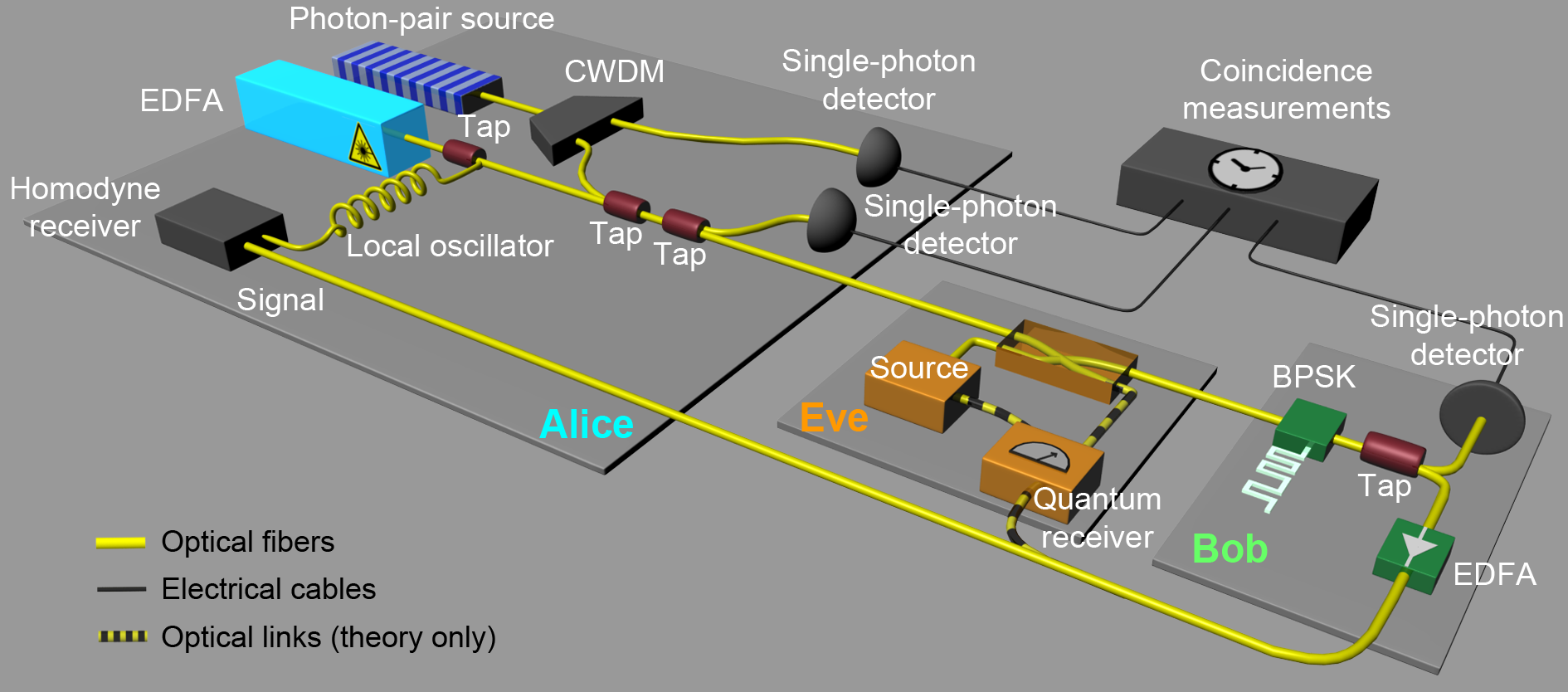}
\caption{\label{FigExp} (color online) FL-QKD's experimental implementation. BPSK: binary phase-shift keying; CWDM: coarse wavelength-division multiplexer; EDFA: erbium-doped fiber amplifier; Tap: beam splitter. Photon-pair source is a cw SPDC.} 
\end{figure*}

Figure~\ref{FigExp} shows our experimental setup, which was designed to realize two main goals:  demonstrating high-SKR FL-QKD operation; and confirming the feasibility of coincidence-based measurement of Eve's injection fraction, $f_E$, down to values such that the upper bound on her Holevo information rate is severely restricted. Using our existing equipment we chose a moderate modulation rate of 100 Mbit/s, with the understanding that it can be easily scaled to $\sim$10-Gbit/s rates with commercially available components.

Alice uses an erbium-doped fiber amplifier (EDFA) followed by a 18-nm ($W$ = 2.2 THz) optical filter centered at 1550\,nm to produce broadband ASE light. A low-brightness portion of the ASE is tapped for transmission to Bob, while the remainder is stored in a fiber delay loop to serve as the LO for homodyne reception in the decoding step.  The SPDC light is generated in a MgO-doped periodically-poled lithium niobate (PPLN) crystal cw-pumped at 780 nm. The idler arm is detected by a WSi superconducting nanowire single-photon detector (SNSPD) immediately after its production. The signal arm of the SPDC, serving as the channel-monitor probe, is mixed with the ASE light using a 98/2 beam splitter. Before Alice sends the ASE-SPDC light to Bob, she taps and directs a small portion ($< 0.1\%$) of it to a WSi SNSPD for coincidence measurement with the idler. We insert a 10\,dB optical attenuator in the Alice-to-Bob channel to simulate the channel loss of a 50\,km fiber link. Upon receiving the signal, Bob encodes messages using a phase modulator that takes the pseudo-random data generated from a bit-error rate (BER) tester and imparts a 100\,Mbit/s BPSK modulation. He then taps and sends a small portion ($< 0.1 \%$) of the modulated light to a WSi SNSPD for coincidence measurement with the idler. An EDFA amplifies the rest of the signal and masks it with strong ASE noise. At Alice's receiver, she combines the returned signal and her stored and delay-matched LO to perform a broadband homodyne measurement to retrieve Bob's message (See Ref.~\cite{SM} for more details).

\begin{figure}[!ht]
\includegraphics[width=3.5in]{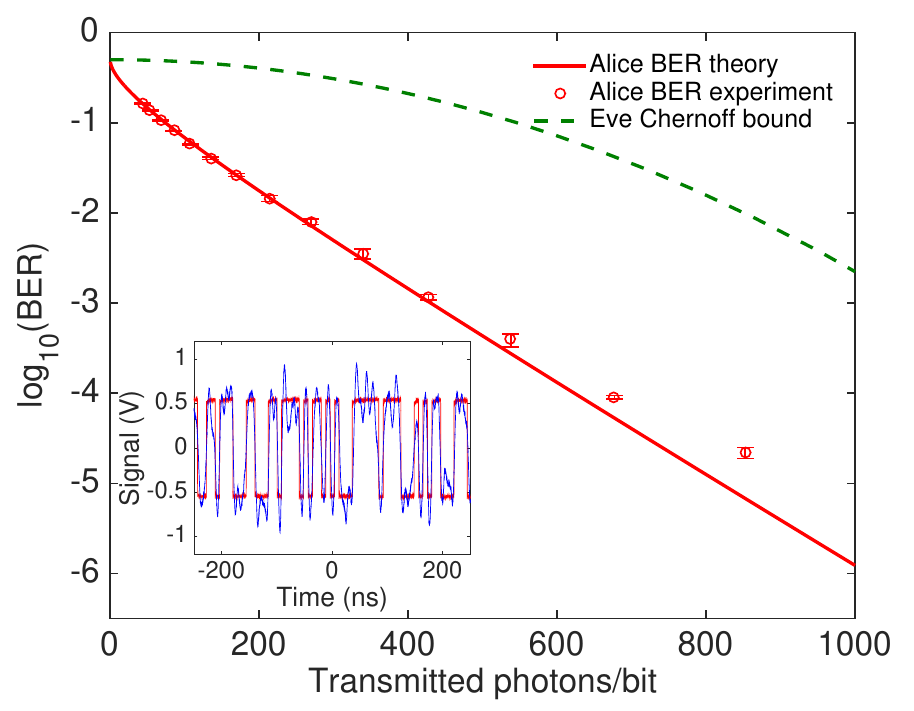}
\caption{\label{FigBER_vs_NS_pBit} (color online) Alice's BER versus her mean number of transmitted photons per bit duration for $W$ = 2.2\,THz source bandwidth and $R$ = 100\,Mbit/s modulation over a channel with 10\,dB loss. Red circles: measured BER with error bars indicating $\pm 1$ standard deviation of 10 consecutive measurements. Solid (red) curve: predicted BER, including all measured experimental imperfections (detailed in Ref.~\cite{SM}). Dashed (green) curve: quantum Chernoff bound on Eve's BER for her optimum passive individual attack. Inset: 50\,bits of homodyne-receiver's output (blue jagged curve) versus Bob's modulated waveform (red square-wave) at a BER level of 10$^{-4}$, clearly showing the high signal-to-noise ratio of Alice's receiver.}
\end{figure}

To estimate Alice and Bob's Shannon information, BER measurements were taken at different source brightness levels, as plotted in Fig.~\ref{FigBER_vs_NS_pBit}, showing good agreement between theory and experiment. At all values of Alice's transmitted source brightness $N_S = {\rm PPB}/WT$, where PPB is her mean number of transmitted photons per bit duration, Alice's BER is far below the quantum Chernoff bound on Eve's BER for her optimum passive individual attack \cite{Shapiro2009}. Experimental instabilities, such as mechanical vibrations, thermal fluctuations, and polarization drifts, cause the BER deviate from theory at a level of $\sim 1\times 10^{-5}$, which has no effect on the SKRs. The inset of Fig.~\ref{FigBER_vs_NS_pBit} overlays 50 bits of the homodyne receiver's real-time output on Bob's scaled modulated message waveform, showing high signal-to-noise ratio for message decoding.  Note that the homodyne measurement noise was dominated by the ASE of Bob's amplifier, which was $\sim$20\,dB above the shot-noise level. 

\begin{figure}[!ht]
\includegraphics[width=3.4in]{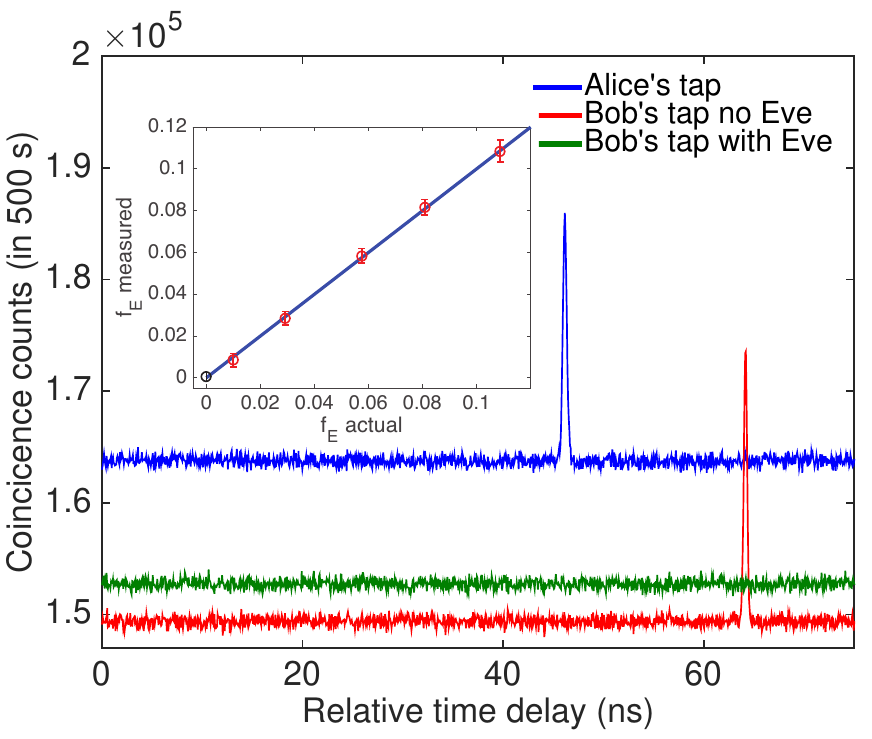}
\caption{\label{Fighist_EveInj} (color online) Photon-coincidence histogram of the channel monitor. Top curve (blue): Alice's tap; Bottom curve (red): Bob's tap without Eve's light injection; Middle curve (green): Bob's tap when Eve injects all the light entering his terminal. Inset: measurements of $f_E$ (red markers) under Eve's intrusion at different injection fractions with error bars indicating $\pm 1$ standard deviation over ten 500-s measurements; and measurement of $f_E$ (black marker) when there is no actual injection with an error bar indicating $\pm 1$ standard deviation over thirty 500-s measurements.}
\end{figure}

We simulated Eve's SPDC-injection attack by feeding broadband light into Bob's terminal via a beam splitter. Figure~\ref{Fighist_EveInj} illustrates FL-QKD's ability to detect intrusion with coincidence-data histogram plots from Alice and Bob's monitors, both referenced to the arrival time of the idler photons. Here, the histograms of Alice's reference tap, Bob's tap without Eve's light injection ($f_E = 0$), and Bob's tap with Eve replacing Alice's light with her own ($f_E=1$) correspond to the top (blue), bottom (red), and middle (green) curves, respectively. Note the absence of a coincidence peak for Bob's tap when $f_E =1$. To demonstrate the effectiveness of Alice and Bob's channel monitoring, we compare the measured $f_E$ value, via Eq.~(\ref{monitor}), with Eve's actual $f_E$ value, as determined by direct power measurements of Alice's and Eve's light entering Bob's terminal.  The inset in Fig.~\ref{Fighist_EveInj} shows excellent agreement between the measured and actual $f_E$ values.

\begin{figure}[ht]
\includegraphics[width=3.4in]{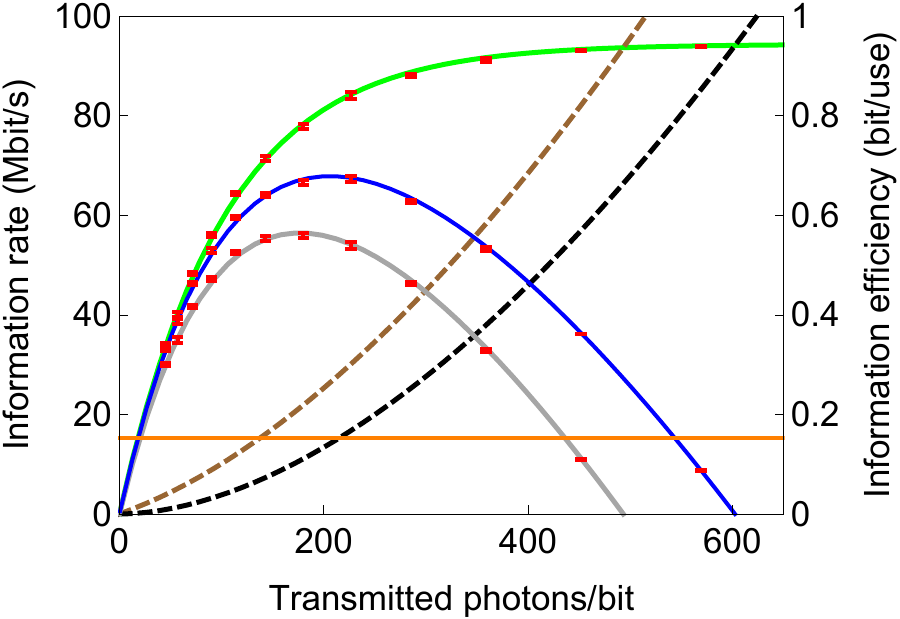}
\caption{\label{Figrate} (color online) Information rates and efficiencies in Mbit/s (left axis) and bits per channel use (right axis) versus Alice's mean number of transmitted photons per bit duration. Solid top curve (green): Alice and Bob's Shannon information; Solid middle curve (blue): SKR against a passive Eve; Solid bottom curve (gray): SKR against the optimum collective attack; Dashed top curve (brown): upper bound on Eve's Holevo information for the optimum collective attack; Dashed bottom curve (black): upper bound on Eve's Holevo information for a passive attack; Horizontal solid line: ultimate limit on SKE in bits per channel use for a 10-dB-loss channel for {\em all} single-mode-per-channel-use one-way QKD protocols \cite{Pirandola2015}.}
\end{figure}

To estimate the SKR, we turned off Eve's injection and measured $f_E = (0.07 \pm 0.2)\%$ from the channel monitors. After adding one standard deviation of measurement uncertainty (other confidence levels are in Ref.~\cite{SM}), we use $f^{\rm UB}_E = 0.27\%$ as an upper bound on Eve's injection fraction. We evaluate Alice and Bob's Shannon information from the BER measurements, Eve's Holevo-information upper bound $\chi^{\rm UB}_E(f^{\rm UB}_E)$ for her optimum collective attack, and $\chi^{\rm UB}_E(0)$ for her passive attack (see Ref.~\cite{SM}). In this SKR estimation, we assume a $\beta =  94\%$ reconciliation efficiency \cite{Richardson2001}. For a passive attack with our 100\,Mbit/s modulation rate, we optimize Alice's mean number of transmitted photons per bit and obtain a peak direct secure-communication efficiency of 0.66\,bits per channel use, yielding a direct secure-communication rate of 66\,Mbit/s. Against Eve's optimum collective attack with the same 100\,Mbit/s modulation rate, we obtain a peak SKE of 0.55\,bits per channel use, after optimizing Alice's mean number of transmitted photons per channel bit, giving rise to an SKR of 55\,Mbit/s. This SKE exceeds the ultimate limit for all one-way single-mode-per-channel-use QKD protocols by 5.6\,dB, see Fig.~\ref{Figrate}. 

In conclusion, we have experimentally demonstrated high-rate FL-QKD using classical-state broadband light augmented by a photon-pair source for channel monitoring. Our experiment's 55\,Mbits/s SKR over a 10-dB-loss channel in an asymptotic regime is a $\sim$50-fold improvement over state-of-the-art QKD for the same channel attenuation \cite{Lucamarini2013,Comandar2014}, and our experiment's SKE of 0.55 bits per channel use is a $\sim$500-fold improvement over that state-of-the-art system's efficiency. A complete finite-size analysis, as done in Refs.~\cite{Lucamarini2013,Comandar2014}, is part of our future study, but we expect that FL-QKD's high SKR would allow it to approach the asymptotic limit in a few minutes.

While pursuing FL-QKD's full security proof against general coherent attacks, we have analyzed a class of such attacks, including an intercept-and-resend attack \cite{Zhuang2016}, and found FL-QKD secure owing to the number-phase uncertainty principle \cite{SM}. From an experimental perspective, we expect to substantially boost FL-QKD's throughput by increasing the modulation rate, optimizing the source brightness, and employing faster NbN SNSPDs to reduce the integration time needed for channel monitoring. Additionally, we plan to implement FL-QKD with installed fibers and dispersion compensation, which we have already done in the current experiment using dispersion-compensating components. With these future developments, FL-QKD points to a viable route to long-distance Gbit/s communication systems with certifiable security. FL-QKD, like prevailing QKD protocols, only allows point-to-point communication, so that extending the protocol to enable secure communication between distributed users in a network would be intriguing. 

\begin{acknowledgments}
We thank Q. Zhao and A. McCaughan for valuable discussions about SNSPDs and E. Wong for generating the 3D experimental schematic. This research was funded by ONR Grant number N00014-13-1-0774, AFOSR Grant number FA9550-14-1-0052, the DARPA Quiness Program through U.S. Army Research Office Grant number W31P4Q-12-1-0019, and DURIP instrumentation Grant number N00014-14-1-0808.
\end{acknowledgments}

\begin{widetext}
\section{Supplemental Material}
\subsection{Experimental details} 
\label{sec:detailed_experimental_setup}

\begin{figure}[ht]
\includegraphics[width=6in]{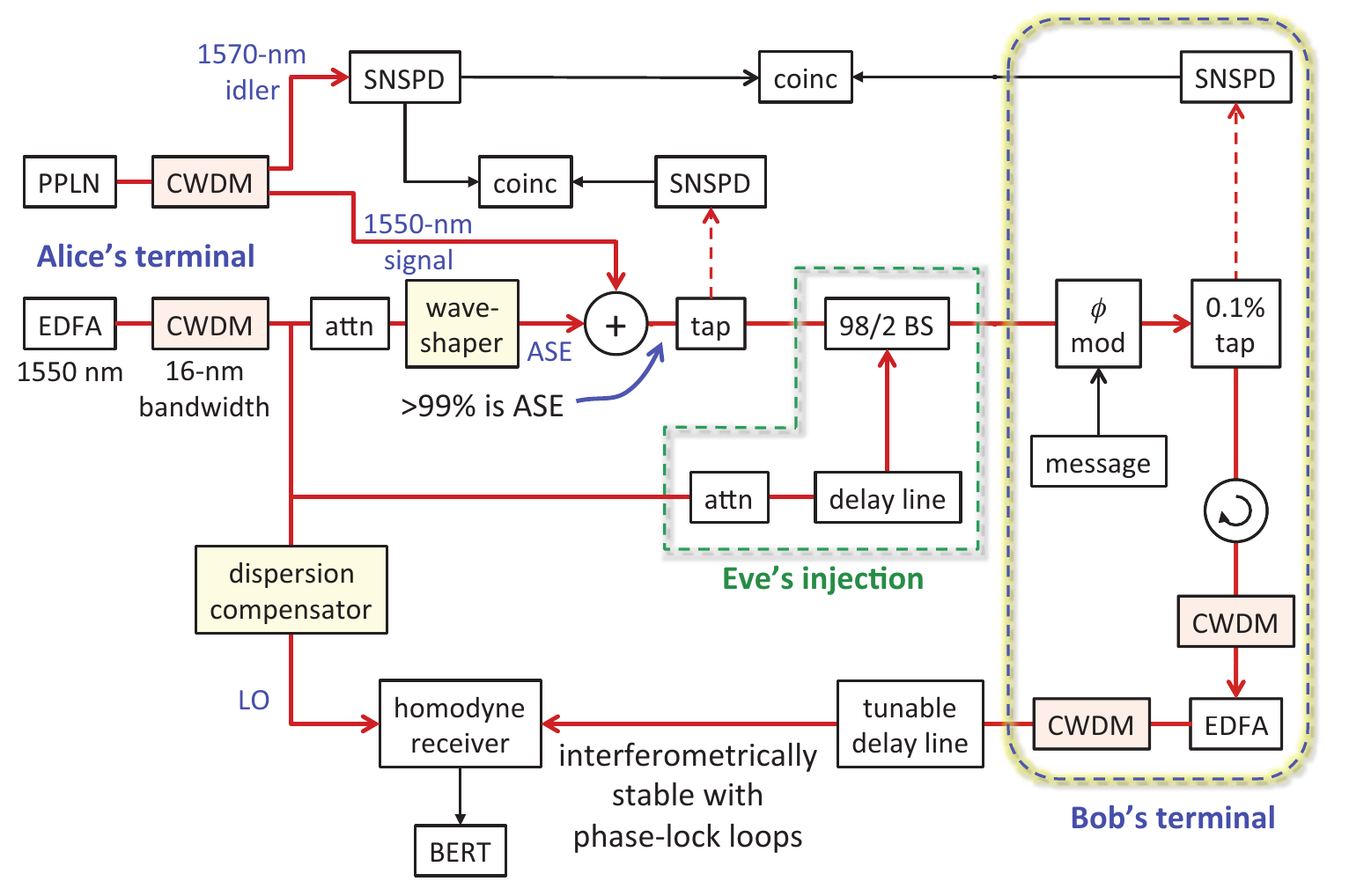}
\caption{\label{FigExpSupp} Experimental schematic. Attn: attenuator; BERT: bit-error rate tester; BPSK: binary phase-shift keying; Coinc: coincidence counting; CWDM: coarse wavelength-division multiplexer; EDFA: erbium-doped fiber amplifier; LO: local oscillator; PPLN: periodically-poled lithium niobate; SNSPD: superconducting nanowire single-photon detector; SPDC: spontaneous parametric down-converter.} 
\end{figure}

A detailed block diagram of our FL-QKD experimental setup is shown in Fig.~\ref{FigExpSupp}. In the experiment, the spontaneous parametric down-converter (SPDC) is a non-degenerate type-0 phase-matched, MgO-doped, periodically-poled lithium niobate (PPLN) crystal that is continuous-wave pumped at 780 nm with its output coupled into a single-mode fiber. A coarse wavelength-division multiplexer (CWDM) separates the SPDC signal and idler into two flat-top 18-nm-wide channels centered at 1550 nm and 1570 nm, respectively. The idler is detected immediately by a WSi superconducting nanowire single-photon detector (SNSPD) with $\sim$80\% detection efficiency and a counting rate of $\sim$2M counts/s in an unsaturated regime. The SPDC signal is used as a probe for quantifying Eve's active intrusion. The amplified spontaneous emission (ASE) noise produced by an erbium-doped fiber amplifier (EDFA) is filtered by a CWDM, whose 18-nm-wide 1550 nm output channel provides the classical broadband light. The classical broadband light is split by a 99:1 beam splitter: the 99\% output serves as the LO for Alice's homodyne receiver to be stored in a fiber spool situated inside Alice's terminal, and the 1\% output is further attenuated by a tunable optical attenuator for transmission to Bob.  

The amplitude and phase of the ASE light for transmission are fine tuned using a WaveShaper (Finisar 1000S) that offers two functionalities. First, the WaveShaper introduces a wavelength-dependent loss on the ASE light to render its spectrum indistinguishable from the spectrum of the SPDC signal. Matching their spectra is critical for the security of the protocol. Second, the WaveShaper applies a wavelength-dependent phase shift on the ASE light to fine tune its dispersion property. In the experiment we find that the WaveShaper is particularly useful for compensating high-order dispersion. The waveshaped ASE light is combined with the SPDC signal on a 98:2 beam splitter: 98\% of the SPDC signal and 2\% of the waveshaped ASE light are sent to Bob. For Alice's reference tap, this combined ASE-SPDC light is tapped ($< 0.1\%$, limited by the deadtime of the SNSPDs) for detection with a second WSi SNSPD, and the rest of the ASE-SPDC light is sent to Bob.  To simulate the channel loss induced by a 50-km fiber link, we insert a 10\,dB optical attenuator in the Alice-to-Bob channel. In the experiment, the tunable optical attenuator is used to adjust the mean number of transmitted ASE photons per bit. At the operating point of $\sim$200 ASE photons per bit, the SPDC signal constitutes $<$0.05\% of the combined transmitted ASE-SPDC light. 

At Bob's terminal, we first employ a CWDM to reject all out-of-band light followed by a circulator to block all back propagating light. We then take the pseudo-random output from a bit-error rate tester (BERT) to encode message bits onto the CWDM-filtered light through a phase modulator at 100-Mbit/s using binary phase-shift keying (BPSK). The polarizer embedded in the phase modulator blocks all unwanted polarizations. To monitor the channel, a small portion ($ < 0.1 \% $, limited by the deadtime of the SNSPDs) is tapped and detected by a WSi SNSPD. We use an EDFA with a measured $\sim$7\,dB weak-input noise figure to amplify and mask the encoded message. A CWDM is installed after the EDFA to reject all out-of-band ASE noise. 

At Alice's receiver, we build a free-space tunable delay line to match the propagation delays incurred by the LO and the light that has undergone the Alice-to-Bob-to-Alice roundtrip. To compensate the dispersion in the fiber link, we pass the LO through $\sim $3\,m of dispersion-compensating fiber. The delay-matched LO and returned signal are mixed on a 50:50 beam splitter prior to their input to a 75-MHz bandwidth balanced receiver (Thorlabs PDB-420C). To compensate the phase drift arising from thermal and mechanical fluctuations, we implement a servo loop to lock the relative phase between the LO and the returned signal. In the servo loop, a lock-in amplifier outputs a 5\,kHz sinusoidal dither that is superimposed on the BPSK modulation. At the balanced receiver, the high-frequency (BPSK modulation rate) amplitude of the output signal is rectified with a power splitter followed by a radio-frequency double-balanced mixer and then fed back to the lock-in amplifier to generate an error signal for the dither. A proportional-integral-derivative controller processes the output of the lock-in amplifier and generates a phase-compensation signal to be added to Bob's phase modulator. The homodyne receiver's output is electrically filtered and then either directed to the BERT for bit-error rate (BER) measurement, or to a wide-band oscilloscope for waveform measurement. 

To simulate Eve's SPDC-injection attack, we add a 98:2 beam splitter (insertion loss included in $\kappa_S$) to the Alice-to-Bob channel. We split off a small amount of ASE light from the LO, introduce a time delay and optical attenuation, and inject the broadband light into Bob's terminal through the 2\% port of the beam splitter. In this way, Eve's injected light has the same spectrum as the light Alice sends to Bob, and a sufficiently long time delay ensures that Eve's injected light and the ASE light Alice sends are not correlated at Bob's terminal.

\subsection{Security analysis} 
\label{sec:detailed_security_analysis}
Detailed security analysis for FL-QKD has been presented in Ref.~\cite{Zhuang2016}. Here, we outline the essential components of that analysis. In both her passive and optimum collective attacks, Eve replaces the lossy Alice-to-Bob fiber with lossless fiber into which she has inserted a beam splitter, and, for analysis purposes, she is granted {\em all} photons in the Bob-to-Alice channel. In her passive attack, Eve captures all photons at the output port of the beam splitter in the Alice-to-Bob channel without injecting any light into that channel. In the SPDC-injection realization of her optimum collective attack, Eve produces quadrature-entangled light from a SPDC and injects the SPDC signal through her input port of the beam splitter in the Alice-to-Bob channel. We assume Eve has a perfect quantum memory to store all captured photons as well as her SPDC idler. At Eve's receiver, she either decodes each bit individually or performs a collective measurement on all photons she possesses. In the individual attack, we use the quantum Chernoff bound to quantify Eve's optimal BER. Eve's capability in the collective attack is quantified by an upper bound on her Holevo information. 

In a passive individual attack, the quantum Chernoff bound for Eve's optimal discrimination strategy is given by
\begin{equation}
{\rm Pr}(e)^{\rm QCB}_E = \frac{1}{2}\exp\left[-4M\kappa(1-\kappa)(1-\kappa_B) N_S^2\right],
\end{equation}
where $\kappa$ is the transmissivity of the Alice-to-Bob channel, $\kappa_B$ is the device loss prior to the EDFA at Bob's terminal, and $N_S$ is Alice's source brightness as measured at the beginning of the Alice-to-Bob channel.

Our lower bound on Alice and Bob's secret-key rate against collective attacks is given by
\begin{equation}
	\Delta R_{AB}^{\rm LB}(f_E) = \left[\beta I_{AB}-\chi_{EB}^{\rm UB}(f_E)\right]R,
\end{equation}
where $R$ is Bob's modulation rate. Our experiment used $R = 100$ Mbit/s, and our secret-key rate calculation assumes $\beta = 0.94$ \cite{Richardson2001}. In FL-QKD, Bob's BPSK modulation leads to a binary signal from Alice's homodyne measurement, so that  $I_{AB}$ is directly determined by Alice's bit-error rate ${\rm Pr}(e)_{\rm Alice}^{\rm hom}$:
\begin{eqnarray}
I_{AB} &=& 1+{\rm Pr}(e)_{\rm Alice}^{\rm hom}\log_2[{\rm Pr}(e)_{\rm Alice}^{\rm hom}]\notag\\
&+&[1-{\rm Pr}(e)_{\rm Alice}^{\rm hom}]\log_2[1-{\rm Pr}(e)_{\rm Alice}^{\rm hom}].
\end{eqnarray}

Alice's theoretical BER is calculated using Eq.~(D8) of Ref.~\cite{Zhuang2016}'s Appendix D augmented by a parameter $\eta <1$ that models experimental imperfections caused by residual dispersion and electronic-filter mismatch, and $\kappa_B<1$ that models device losses in Bob's terminal prior to his EDFA. Furthermore, $\kappa_S$ in Eq.~(D8) of Ref.~\cite{Zhuang2016}'s Appendix D refers to the channel transmissivity seen by Alice and Bob, i.e., Bob's received photon flux divided by Alice's transmitted photon flux. To relate $\kappa_S$ to the actual channel transmissivity $\kappa$ (determined by the loss induced by the optical attenuator), we let $\kappa_S$ in Eq.~(D8) of Ref.~\cite{Zhuang2016}'s Appendix D be $\kappa/(1-f_E)$ and obtain
\begin{eqnarray}
\label{eqBER}
	{\rm Pr}(e)_{\rm Alice}^{\rm hom} &=& Q\!\left(\sqrt{2M\kappa\eta(1-\kappa_B) N_S G_B/N_B}\right)\\[.05in]
	&=&Q\!\left(\sqrt{2M\kappa\eta(1-\kappa_B) N_S /\gamma}\right),
\end{eqnarray}
where
\begin{equation}
	Q(x) = \int_x^\infty dt\, \frac{e^{-t^2/2}}{\sqrt{2\pi}}.
\end{equation}
In Eq.~(\ref{eqBER}), $N_S$ is the source brightness, $\kappa = 0.1$ is the one-way transmissivity, $M = 2.0\times 10^4$ is the number of modes per bit, $G_B = 3.8\times 10^3$ is the EDFA's gain, $N_B = 9.7\times 10^3$\,photons/s-Hz is the brightness of the EDFA's ASE output, and $\eta \sim 0.9$, $\kappa_B = 0.71$ are obtained by experimental calibration.  The EDFA's noise figure $10\log_{10}(\gamma)+3$, where $\gamma \equiv N_B/G_B$,  was measured to be $\sim$7\,dB with a weak-signal input. The source brightness is calculated using 
\begin{equation}
	N_S = \frac{P}{\hbar\omega_0 W},
\end{equation}
where $P$ is the measured signal power at the output of Alice's terminal, $\hbar\omega_0 \approx 1.28 \times 10^{-19}$\,J is the signal wavelength's photon energy, and $W= M\times R = 2.2$\,THz (18\,nm) is the signal bandwidth. 

We next derive our upper bound on Eve's Holevo information. We consider Eve's optimum collective attack, in which she employs a broadband SPDC source to produce quadrature-entangled signal and idler beams and uses a beam splitter to inject the signal light into Bob's terminal. The spectra of Alice's signal and the signal arm of Eve's SPDC source are well matched, because any out-of-band photons are filtered out by the CWDM that precedes Bob's EDFA. Our upper bound on Eve's Holevo information rate is given by \cite{Zhuang2016}
\begin{equation}
	\chi_{EB}^{\rm UB}(f_E) = \min \!\left\{S\!\left[\boldsymbol{\rho}_E^{\rm Gauss}(f_E)\right]-\frac{1}{2}\sum_{k = 0}^1 S\!\left[\boldsymbol{\rho}_E^{(k)}(f_E)\right],1\right\}.
\end{equation}
Here: $\boldsymbol{\rho}_E^{(k)}$ is Eve's conditional density operator---given the value, $k$, of Bob's bit---for the light at her disposal. It is an $M$-fold tensor product of zero-mean, 3-mode Gaussian states all with the same Wigner covariance matrix $\boldsymbol{\Lambda}_E^{(k)(f_E)}$;   and $\boldsymbol{\rho}_E^{\rm Gauss}(f_E)$ is an $M$-fold tensor product of zero-mean, 3-mode Gaussian states all with the same Wigner covariance matrix $\boldsymbol{\Lambda}_E(f_E) = \sum_{k = 0}^1 \boldsymbol{\Lambda}_E^{(k)}(f_E)/2$; and $S(\cdot)$ denotes von Neumann entropy. 

Eve's conditional von Neumann entropy satisfies
\begin{equation}
	S\!\left[\boldsymbol{\rho}_E^{(k)}(f_E)\right] = MS\!\left[\rho_E^{(k)}(f_E)\right],
\end{equation}
where $\rho_E^{(k)}$ is a zero-mean, three-mode Gaussian state with Wigner covariance function
\begin{eqnarray}
	\lefteqn{\Lambda_{E}^{(k)}(f_E) = } \nonumber \\[.05in]
	&&\frac{1}{4}\left[ \begin{array}{cccccc}
2N_{AB}^{\rm act}(f_E)+1 & 0 & -C_{IA}^{\rm act}(f_E) & 0 & (-1)^k C_{AB}^{\rm act}(f_E) & 0 \\[.05in]
0 & 2N_{AB}^{\rm act}(f_E)+1 & 0 & C_{IA}^{\rm act}(f_E) & 0 & (-1)^k C_{AB}^{\rm act}(f_E) \\[.05in]
-C_{IA}^{\rm act}(f_E) & 0 & 2N_E(f_E)+1 & 0 & (-1)^k C_{IB}^{\rm act}(f_E) & 0 \\[.05in]
0 & C_{IA}^{\rm act}(f_E) & 0 & 2N_E(f_E)+1 & 0 & (-1)^{k+1}C_{IB}^{\rm act}(f_E) \\[.05in]
(-1)^k C_{AB}^{\rm act}(f_E) & 0 & (-1)^{k} C_{IB}^{\rm act}(f_E) & 0 & 2N_{BA}^{\rm act}(f_E)+1 & 0\\[.05in]
0 & (-1)^k C_{AB}^{\rm act}(f_E) & 0 & (-1)^{k+1} C_{IB}^{\rm act}(f_E) & 0 & 2N_{BA}^{\rm act}(f_E)+1\\\end{array}
\right],
\end{eqnarray}
where
\begin{eqnarray}
\label{eqN_AB} N_{AB}^{\rm act}(f_E) &=& \langle \hat{e}^{(2)\dag}_{I_m} \hat{e}^{(2)}_{I_m}\rangle = (1-\kappa)N_S+\kappa N_E(f_E)\\[.05in]
\label{eqC_AI} C_{IA}^{\rm act}(f_E) &=& \langle \hat{e}^{(1)}_{I_m} \hat{e}^{(2)}_{I_m}\rangle = 2\sqrt{\kappa N_E(f_E)[N_E(f_E)+1]}\\[.05in]
\label{eqC_AB} C_{AB}^{\rm act}(f_E) &=& \langle \hat{e}^{(2)\dag}_{I_m} \hat{a}_{B_m}\rangle = 2\sqrt{G_B(1-\kappa_B)\kappa(1-\kappa)}[N_S-N_E(f_E)]\\[.05in]
\label{eqC_IB} C_{IB}^{\rm act}(f_E) &=& \langle \hat{e}^{(1)}_{I_m} \hat{a}_{B_m}\rangle = 2\sqrt{G_B(1-\kappa_B)(1-\kappa)N_E(f_E)[N_E(f_E)+1]}\\[.05in]
\label{eqN_BA} N_{BA}^{\rm act}(f_E) &=& \langle \hat{a}_{B_m}^\dag \hat{a}_{B_m}\rangle = G_B(1-\kappa_B)\left[\kappa N_S + (1-\kappa)N_E(f_E) \right]+N_B.
\end{eqnarray}
In the preceding expressions: $N_{AB}^{\rm act}(f_E)$ is the brightness of the light Eve tapped from the Alice-to-Bob channel; $N_{BA}^{\rm act}(f_E)$ is the brightness of the light in the Bob-to-Alice channel; $N_E(f_E)$ is the brightness of Eve's SPDC signal; $C_{AB}^{\rm act}(f_E)$ is the quadrature correlation between the light Eve tapped from the Alice-to-Bob channel and the light in the Bob-to-Alice channel; $C_{IA}^{\rm act}(f_E)$ is the quadrature correlation between the light Eve tapped from the Alice-to-Bob channel and her idler; and $C_{IB}^{\rm act}(f_E)$ is the quadrature correlation between the light in the Bob-to-Alice channel and Eve's idler. Equations~(\ref{eqN_AB})--(\ref{eqN_BA}) are obtained using the annihilation operators $\hat{e}^{(1)}_{I_m}$, $\hat{e}^{(2)}_{I_m}$, and $\hat{a}_{B_m}$ defined in Eqs.~(C58), (C59), and (A8) of Ref.~\cite{Zhuang2016}'s Appendices. The annihilation operator $\hat{e}^{(1)}_{I_m}$ is associated with Eve's $m$-th locally-stored idler mode, $\hat{e}^{(2)}_{I_m}$ is associated with the $m$-th mode Eve captures from the Alice-to-Bob channel, and $\hat{a}_{B_m}$ is associated with the $m$-th returned signal mode from Bob.  Alice and Bob's channel monitoring provides a calibration-free measurement of Eve's injection fraction $f_E$ \cite{Zhuang2016}, from which they can compute the brightness of Eve's SPDC signal light as follows:
\begin{equation}
N_E(f_E) = \frac{\kappa N_S f_E}{(1-\kappa)(1-f_E)}.
\end{equation}

Eve's unconditional Wigner covariance matrix is $6M \times 6M$ block diagonal with $6 \times 6$ identical blocks given by
\begin{eqnarray}
	\lefteqn{\Lambda_{E}(f_E) = \sum_{k=0}^1 \Lambda^{(k)}_{E}(f_E)/2 = } \\[.05in]
	&& \frac{1}{4}\left[ \begin{array}{cccccc}
2N_{AB}^{\rm act}(f_E)+1 & 0 & -C_{IA}^{\rm act}(f_E) & 0 & 0 & 0  \\[.05in]
0 & 2N_{AB}^{\rm act}(f_E)+1 & 0 & C_{IA}^{\rm act}(f_E) & 0 & 0  \\[.05in]
-C_{IA}^{\rm act}(f_E) & 0 & 2N_E(f_E)+1 & 0 & 0 & 0  \\[.05in]
0 & C_{IA}^{\rm act}(f_E) & 0 & 2N_E(f_E)+1 & 0 & 0  \\[.05in]
0 & 0 & 0 & 0 & 2N_{BA}^{\rm act}(f_E)+1 & 0 \\[.05in]
0 & 0 & 0 & 0 & 0 & 2N_{BA}^{\rm act}(f_E)+1\\\end{array}
\right].
\end{eqnarray}

With $\Lambda_{E}^{(k)}(f_E)$ and $\Lambda_{E}(f_E)$ in hand, one can readily evaluate $S\!\left[\boldsymbol{\rho}_E^{\rm Gauss}(f_E)\right]$ and $S\!\left[\boldsymbol{\rho}_E^{(k)}(f_E)\right]$ using symplectic decomposition \cite{Pirandola2008_QCB}.

FL-QKD's security analysis has so far been established against collective attacks in which Eve interacts individually with each frequency mode but is allowed to perform an optimum joint measurement over all modes. A full security proof against general coherent attack is under development, but this does not preclude us from analyzing FL-QKD's security against specific coherent attacks, one of which is the intercept-and-resend (I\&R) attack. In such an attack, Eve measures the timing of each photon from Alice and endeavors to elude the channel monitor by producing a quantum signal that mimics the measured photon statistics. In doing so, Eve interacts coherently with {\em all} frequency modes. I\&R attacks have been given an appreciable amount of consideration in Ref.~\cite{Zhuang2016}, and here let us formulate the following I\&R attack. Eve first takes the light from Alice, performs a quantum non-demolition (QND) photon-number measurement on every temporal mode, and stores the output light from that measurement for use as a reference. To elude the channel monitor, Eve needs to ensure that: (1) she sends a vacuum state $|0\rangle$ into Bob's terminal when the measured temporal mode contains no photon; and (2) she sends the Fock state $|1\rangle$ into Bob's terminal when the measured temporal mode contains a photon. This is Eve's only strategy to stay undetected because violation of the former increases the time-shifted coincidence rate $\widetilde{C}_{IB}$ while violation of the latter reduces the time-aligned coincidence rate $C_{IB}$, either of which leads to $f_E \neq 0$ being measured. At this juncture, let us consider the amount of information Eve can acquire in either case. In the former case, a vacuum mode obviously carries no information. In the latter case, $|1\rangle$'s phase is completely undefined because its photon number is well defined---a consequence of the number-phase uncertainty principle \cite{Carruthers1965}. In other words, phase modulation on $|1\rangle$ leaves the state unchanged, viz. $e^{i\theta}|1\rangle = |1\rangle$ insofar as any measurement on that state is concerned. Therefore, there does not exist a phase reference for $|1\rangle$ that allows for effectively decoding of Bob's phase modulation. 
We thus conclude that the I\&R attack offers Eve no information. The above argument is consistent with FL-QKD's security analysis \cite{Zhuang2016}, which proves that $f_E = 0$ indicates Eve's gaining no information in her {\em active} attacks. Going one step forward, let us analyze how the I\&R attack affects Alice's BER. First note that Eve's QND timing measurements destroy the phase coherence between her retained photons and Alice's local oscillator. In addition, we learned that Eve is unable to decode Bob's phase modulation. Consequently, Eve's encoding on the retained photons yields a 50\% BER at Alice's terminal, leaving her I\&R attack immediately detectable.

\subsection{Secret-key rates at different confidence levels} 
\label{sec:secret_key_rates_at_different_confidence_levels}
In the Letter, the reported SKR of 55\,Mbit/s is obtained by assuming that Eve's injection fraction equals $f_E^{\rm UB}$, the experimentally-determined injection fraction $f_E$ plus one measurement standard deviation ($\sigma$).  SKRs at higher confidence levels, i.e., adding more standard deviations to the experimental $f_E$ value, can also be derived. A key feature of FL-QKD is that its SKR can be optimized over source brightness.  So, we optimize SKR over source brightness at $2 \sigma$, $3 \sigma$, $4\sigma$, and $5\sigma$ confidence levels and obtain the results in Table~1. It is notable that the SKRs at higher confidence levels do not degrade much by virtue of the optimization.

\begin{table}[H]
\begin{center}
    \begin{tabular}{ | l | l |}
    \hline
    Confidence level & Secret-key rate  \\ 
    \hline\hline
    $1 \sigma$ & 55\,Mbit/s \\
    \hline
    $2 \sigma$ & 49\,Mbit/s \\
    \hline
    $3 \sigma$ & 43\,Mbit/s \\
    \hline
    $4 \sigma$ & 38\,Mbit/s \\
    \hline
    $5 \sigma$ & 34\,Mbit/s \\
    \hline

    \end{tabular}
\end{center}
\caption{SKRs at various confidence levels.} 
\end{table}
\end{widetext}

\end{document}